# Visible Emission Spectroscopy on Highly Charged Tungsten Ions in LHD II – Evaluation of Tungsten Ion Temperature -

K. Fujii[1], Y. Takahashi[1], Y. Nakai[1], D. Kato[2,3], M. Goto[2,3], S. Morita[2,3], M. Hasuo[1] and LHD Experiment Group[2]

[1]Department of Mechanical Engineering and Science, Graduate School of Engineering, Kyoto University, Kyoto 615-8540, Japan

[2]National Institute for Fusion Science, Toki 509-5292, Japan

[3]Dept. of Fusion Science, The Graduate University for Advanced Studies, Toki 509-5292, Japan

fujii@me.kyoto-u.ac.jp

**Abstract**

We demonstrated a polarization-resolved high resolution spectroscopy of a visible emission line of highly charged tungsten ions ($\lambda_0$ = 668.899 nm, M. Shinohara et al, Phys. Scr., Submitted) for the Large Helical Device (LHD) plasma, where the tungsten ions were introduced by a pellet injection. Its spectral profile shows broadening and polarization dependence, which are attributed to the Doppler and Zeeman effects, respectively. The tungsten ion temperature was evaluated for the first time from the broadening of visible the emission line, with its emission location determined by the Abel inversion of the chord-integrated emission intensities observed with multiple chords. The tungsten ion temperature was found to be close to the helium-like argon ion temperature, which is used as an ion temperature monitor in LHD.

# 1. Introduction

The ion temperature in the core region of the magnetic plasma confinement devices has been widely evaluated by the charge exchange spectroscopy [1]. In the charge exchange spectroscopy, the emission followed by the charge exchange reaction between fully stripped impurity ions in the plasma and high energy neutral beam injected into the plasma for heating. However in ITER, it will be difficult to implement the charge exchange signal from the core region because of the large attenuation of the neutral beam due to the high density and large diameter plasma [2,3].

For achieving the core temperature measurement in ITER, the passive spectroscopy of the X-ray emission lines of the highly charged tungsten (W) ions has been planned. Tungsten ions will have bounded electrons even in the center of the ITER plasma with $T_e$ of 10-20 keV. The spectral line profile observed by the X-ray crystal spectrometer will give information of the ion temperature and flow velocity. The spectral profile observation of the tungsten X-ray emission lines has been demonstrated by Nakano et al [4]

Highly charged tungsten ions also emit some visible lines mainly due to the magnetic dipole transition. An emission measurement in near-ultraviolet region in magnetic devices has been started recently by Kato et al in Large Helical Device (LHD) [5]. They have observed the spatial distribution of the emission intensity of $W^{26+}$ $[4d^{10}4f_{5/2}4f_{7/2}]_5 \rightarrow [4d^{10}4f_{5/2}^2]_4$ transition (central wavelength: 389.4 nm) in LHD plasma. However, the ion temperature measurement with visible emission lines of highly charged tungsten ions has not been reported.

In this paper, we demonstrated the first evaluation of the ion temperature based on the visible emission line of the highly charged tungsten ions. Due to the restriction of the instruments, we observed an emission line with the central wavelength of $\lambda_0 = 668.90$ nm (improved to be 668.899 nm in this paper), which has been found by our group in the other paper [6]. Since the Zeeman effect modifies the spectral line profile significantly, we resolved the polarization of the emission line. Although the charge state of this emission line is still unknown, we also estimated it in this paper by comparing the emission location with the theoretical calculation.

## 2 Experiment

LHD is a heliotron device, in which a high temperature hydrogen plasma is magnetically confined by a pair of superconducting helical coils. A poloidal cross section of LHD is illustrated in Fig.1 (a). The closed magnetic flux surfaces are shown by ellipsoidal curves. The normalized radius $\rho$ is indicated in the figure, which is a measure of the closed magnetic flux surface; $\rho = 0$ is assigned to the plasma center and $\rho = 1$ is assigned to the last closed flux surface (LCFS). At the plasma center, the magnetic field strength is 2.64 T and it points in the toroidal direction. Outside the LCFS, there is a thick layer illustrated in gray in Fig.1. This layer is called as "ergodic layer", which is an aggregate of open magnetic field lines. These magnetic field lines are connected to divertor plates. The ergodic layer has a complex shape reflecting the three-dimensional helical divertor structure of LHD.

The divertor plates and the first walls of LHD are made of carbon composite and stainless steel (SUS316L), respectively, and no tungsten is used. For the study of the tungsten transport in the plasma, tungsten is injected as a small pellet. The details of the pellet injection system are described in ref. 6 and 7.

A hydrogen plasma is generated in LHD at $t = 3.0$ s and a tungsten pellet is injected into the plasma at $t = 3.85$ s. The temporal evolutions of $T_e$ and electron density $n_e$ at the plasma center measured by the Thomson scattering [8] are shown in Fig.2 (a). After the injection of the pellet, $T_e$ decreases due to the radiation loss by the highly charged tungsten ions, while $n_e$ increases. The temporal evolution of helium-like argon ion temperature ($T_{Ar}^{16+}$) measured by a crystal X-ray spectrometer [9] is also shown in the figure by a blue curve. Although the measured $T_{Ar}^{16+}$ is the result of the spatial average along the line of sight, the dominant emission location of the $T_{Ar}^{16+}$ line is monitored by the X-ray imaging crystal spectrometer (XICS) [10]. The radial distributions of $T_e$ and $n_e$ at $t = 3.83, 3.93, 4.18$, and $4.33$ s are shown in Fig.2 (b) and (c), respectively. $T_e$ and $n_e$ have steep gradients in the ergodic layer and these values at the divertor region are more than $10^2$ times smaller than those on the LCFS [11].

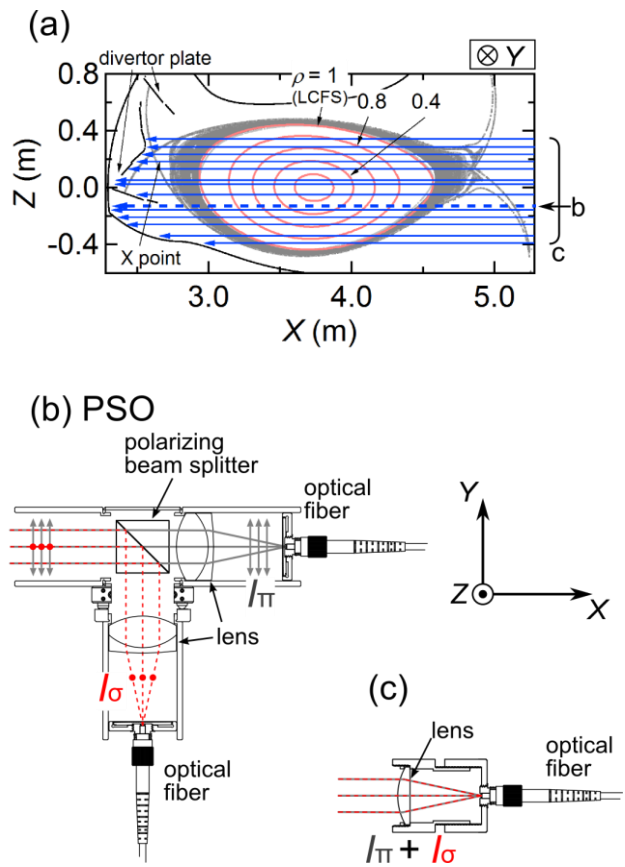

Fig.1(a) A poloidal cross-section of LHD. The closed magnetic flux surfaces and the ergodic layer are shown by red curves and gray dots, respectively. The horizontal dotted and solid arrows indicate the LOSs with and without the polarization resolved measurement, respectively. (b) and (c) Schematic illustrations of the emission collection systems with and without the polarization-resolved optics, respectively.

The lines of sight (LOS) for the tungsten emission measurement are shown in Fig.1 (a) by solid and dotted arrows. The diameter of the LOS at the plasma is roughly 4 cm. For the LOS represented by the dotted arrow, a polarization-resolved optics (PSO, Fig.1(b)) is attached roughly 4 m apart from the plasma. By the PSO, the emission with parallel and perpendicular linear polarization to the magnetic field line at the plasma center are resolved and separately focused on edges of two optical fibers. For the LOSs represented by the solid arrows, the emission was simply focused on an edge of the optical fiber without polarization resolution (Fig.1(c)). The emissions collected for these LOSs are transferred for roughly 50 m by the optical fibers to the entrance slit of the high-throughput and high-resolution spectrometer which we have originally developed for the high dynamic-range Balmer-α spectroscopy [12]. The observable wavelength range of the spectrometer is 663 ~ 672 nm and its instrumental function is well approximated by a Gauss function with its full-width at half-maximum ($\Delta\lambda_{inst}$) of 0.020 nm. The exposure time and the frame rate are 50 ms and 20 Hz, respectively.

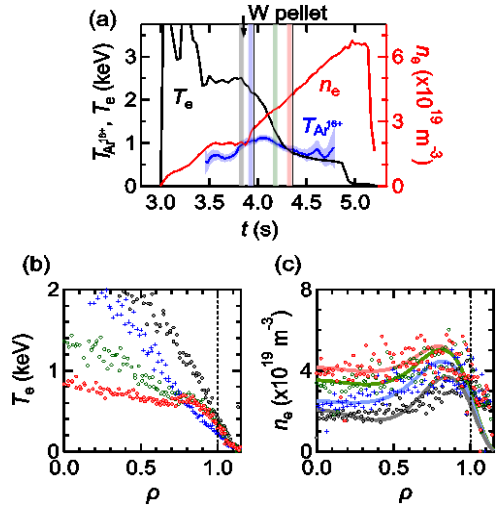

Fig.2 (a) Temporal evolutions of $T_e$, and $n_e$ at the plasma center and the spatially averaged $T_{Ar}^{16+}$. The pellet injection time is indicated by a vertical arrow. The thickness of the blue curve shows the measurement uncertainty of $T_{Ar}^{16+}$. (b) and (c) represent the radial distributions of $T_e$ and $n_e$, respectively, measured at $t$ = 3.83 s (gray circles), 3.93 s (blue circles), 4.18 s (green circles), and 4.33 s (red circles). The smoothed lines for the measured $n_e$ in (c) are eye guides.

## 3 Results

The spectra observed for all the LOSs before ($t = 3.83$ s) and after ($t = 3.93$, 4.18 and 4.33 s) the pellet injection are shown in Fig.3(a) as two dimensional images; the horizontal and vertical axes are the wavelength of light and the height of the LOSs $Z$, respectively, and the intensities of the emission are represented by false color. We note that the continuum emission intensities, which are evaluated later, were subtracted from the shown image for the sake of clarity. In Fig.3(b), the spectra observed for six LOSs at $t = 3.83$ and 4.33 s are plotted with black and red curves, respectively, with vertical offsets.

The emission line of the highly charged tungsten ion having the central wavelength of $\lambda_0 \sim 668.90$ nm is detected after the injection (indicated by vertical a red arrow in Fig.3(b)). Emission lines of argon ions, iron ions and hydrogen molecules are also observed in the wavelength range, the central wavelengths of which are indicated by blue, green and gray vertical bars, respectively. The spectral widths of the these emission lines are close to the instrumental width of the spectrometer (indicated in Fig.3(b) by the interval of two vertical bars pair), while the tungsten emission line shows a significantly broad profile. The spatial distributions of the intensities of argon, iron and hydrogen molecule emission lines are large around $Z \sim 0.2$ m, which corresponds to the height of the inner X-point of the plasma, and around $Z \sim 0.0$ m, which corresponds to the height of the divertor plate (see Fig.1(a)). The spatial distributions of these lines show little temporal changes. On the other hand, the spatial distribution of the tungsten emission line and its temporal evolution are much different; it is a hollow profile at $t = 3.93$ s and becomes a peaked one at $t = 4.33$ s.

The polarization-resolved spectra observed before ($t = 3.83$) and after ($t = 4.33$ s) the injection of the tungsten pellet are shown in Fig.4(a). The gray solid and red open squares show $\pi$- and $\sigma$-components of the emission spectra, $I_\pi$ and $I_\sigma$, respectively. We plot the intensity difference between the two polarization components, $I_\pi - I_\sigma$, in Fig.4 (b). A polarization dependence is clearly seen in the $\lambda_0 \sim 668.90$ emission line observed at $t = 4.33$ s.

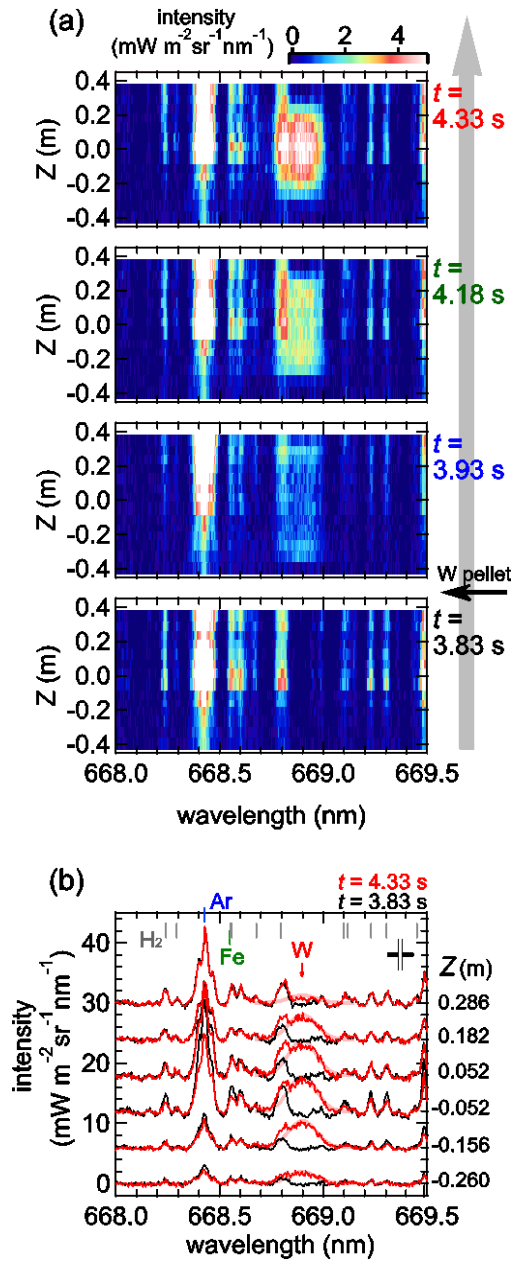

Fig.3 (a) The spectra observed before ($t = 3.80$ s) and after the pellet injection ($t = 3.93$, 4.18 and 4.33 s) as a function of the LOS height ($Z$) and wavelength. The intensity is shown by the false color. The scale is shown on the top-right corner of the figure. The continuum emission intensities are subtracted from the plotted spectra. (b) Observed spectra at $t = 3.83$ s (black curves) and $t = 4.33$ s (red curves) for six LOSs, the height of which are indicated in the right. The instrumental width $\Delta\lambda_{\mathrm{inst}}$ is indicated by the interval of the black vertical bar pair in the top-right. Central wavelengths of the argon ions, iron ions, and hydrogen molecules are indicated by the blue, green, and gray vertical bars, respectively. The fit results for the spectra observed at $t = 4.30$ s are plotted by the bold curves.

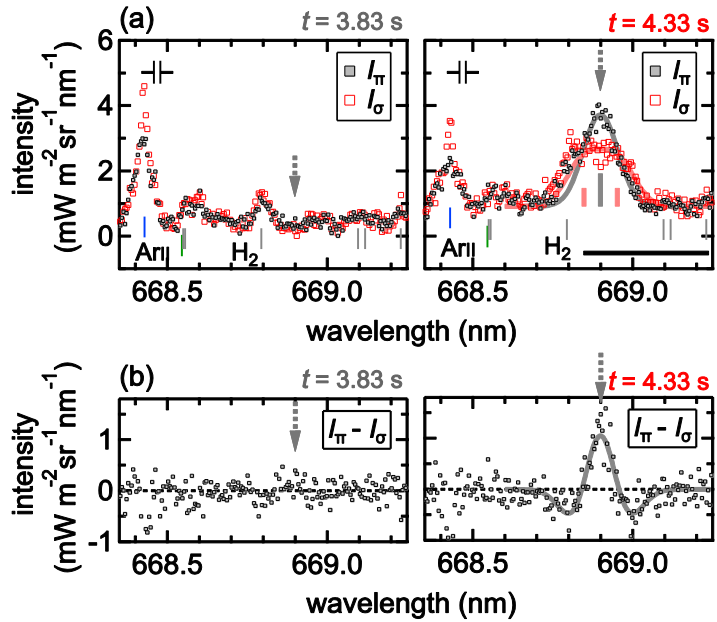

Fig.4 (a) Polarization resolved spectra observed before ($t$ = 3.83 s, left figure) and after the pellet injection ($t$ = 4.33 s, right figure). $I_\pi$ and $I_\sigma$ are plotted by the gray solid and red open squares, respectively. The instrumental width is indicated by the interval of the black vertical bar pair in the top-left of the figure. Central wavelengths of the argon ions, iron ions, and hydrogen molecules are indicated by the blue, green, and gray vertical bars, respectively. (b) The intensity difference between $I_\pi$ and $I_\sigma$. Bold curves in (a) and (b) are fit results for the spectra observed at $t$ = 4.33 s taking the normal Zeeman effect and Doppler effect into account. Thick vertical bars just below the tungsten emission line show the evaluated central wavelengths of the π-component and two σ-components. The horizontal black bar indicates the wavelength region used for the fit.

*3.1 Evaluation of tungsten ion temperature*

The polarization dependence and the broadening of the tungsten emission line can be attributed to the Zeeman effect due to the confinement magnetic field of LHD and the Doppler effect due to the thermal motion of the tungsten ions as well as the instrumental broadening, respectively. Although all information about the emission line, such as a kind of the transition, Landé's *g*-factors of the upper and lower states, are unknown, the observed $I_\pi$ - $I_\sigma$ spectrum suggests a very simple Zeeman profiles originated from the normal Zeeman effect of an electric dipole transition; the π-component does not split and the σ-component splits into two with the shift of $\delta\lambda$ [nm] from the π-component. $I_\pi$ and $I_\sigma$ are expressed as a single and double Gauss functions, respectively, as

$$I_\pi = \frac{I_{cntn}}{2} + \frac{I_W}{2\sqrt{\pi}\Delta\lambda} \exp\left[-\frac{(\lambda-\lambda_0)^2}{\Delta\lambda^2}\right]$$
$$I_\sigma = \frac{I_{cntn}}{2} + \frac{I_W}{4\sqrt{\pi}\Delta\lambda} \exp\left[-\frac{(\lambda-\lambda_0-\delta\lambda)^2}{\Delta\lambda^2}\right] + \frac{I_0}{4\sqrt{\pi}\Delta\lambda} \exp\left[-\frac{(\lambda-\lambda_0+\delta\lambda)^2}{\Delta\lambda^2}\right] \quad (1)$$

with the 1/*e* width of the broadening $\Delta\lambda$ [nm]

$$\Delta\lambda^2 = \lambda_0^2 \frac{2kT_W}{M_W c^2} + 4\ln 2 \Delta\lambda_{inst}^2 \quad (2)$$

where *k* [J/eV] and $M_W$ [kg] are Boltzmann's constant and the mass of a tungsten ion. $I_{cntn}$ [mW m$^{-2}$sr$^{-1}$nm$^{-1}$] is the continuum intensity integrated along the LOS. $I_W$ [mW m$^{-2}$sr$^{-1}$] and $T_W$ [eV] are the LOS-integrated emission line intensity and the temperature of the tungsten ions. The coefficient 4 ln 2 in Eq. (2) is a conversion factor from the FWHM to the 1/*e* width. We fit both the σ- and π-components of the spectrum simultaneously with Eq. (1) in the wavelength range indicated by a horizontal black bar in Fig.4(a). Since a hydrogen molecular line at $\lambda$ = 668.7958 nm [13] contaminates in the shorter wavelength side of the emission line, this region is excluded from the fit. The fit results are shown in Fig.4 by bold curves. The observed spectra are well represented by Eq. (1). Values of $\lambda_0$, $\Delta\lambda$, and $\delta\lambda$ are determined from the polarization resolved spectrum to be 668.899 nm, 0.076 ± 0.003 nm, and 0.052 ± 0.002 nm, respectively. The errors are estimated from the covariant matrix used in the fitting procedure.

From $\Delta\lambda$ evaluated from the polarization resolved spectrum and Eq. (2), the values of $T_W$ are determined. The results are shown in Fig.5 (a) by red circles. The uncertainty of the evaluation is shown by the error bars. The uncertainty at *t* < 4.0 s and *t* > 4.7 s are large because of the small emission intensities. $T_W$ gradually decreases from 1.5 to 1.0 keV.

Since the ion temperature usually has spatial variation in the plasma, the emission location should be determined.

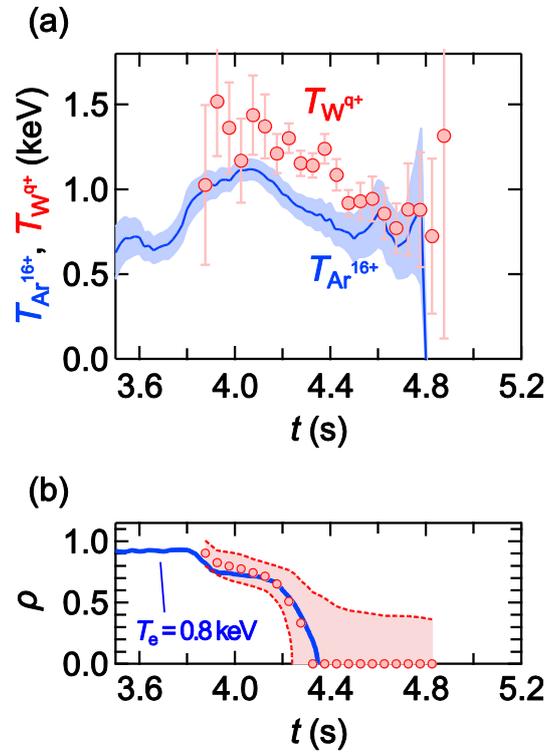

Fig.5 (a) (red circles) Temporal evolutions of $T_W$ estimated from the polarization resolved spectra. (blue curves) The helium-like argon ion temperature measured by the X-ray crystal spectrometer. Its uncertainty is represented by the thickness of the curve. (b) (red circles) The peak position of the emissivity of the tungsten line and (red solid curves) its emission region which is defined by the width of the emissivity. (blue curve) $T_e = 0.8$ keV position.

*3.2 Evaluation of the emission location*

For the other spectra observed without the polarization resolution, we fit them with $I_\pi + I_\sigma$ of Eq. (1), with the adjustable parameter of $I_w$ and $I_{cntn}$, and fixed $\Delta\lambda$, $\delta\lambda$ and $\lambda_0$; the values of $\Delta\lambda$, $\delta\lambda$ and $\lambda_0$ are determined from the fit result of the polarization resolved spectrum observed in the same frame. Although $\Delta\lambda$ and $\delta\lambda$ may depend on the emission location in principle, we confirm that it makes little impact on the evaluation of $I_{cntn}$ and $I_W$. The fit results for the spectra without the polarization resolution are shown by solid curves in Fig.3 (b). These spectra are also well represented. $I_W$ evaluated from the spectra observed at $t = $ 3.93, 4.18, and 4.33 s are shown in Fig.6 by red markers as a function of Z.

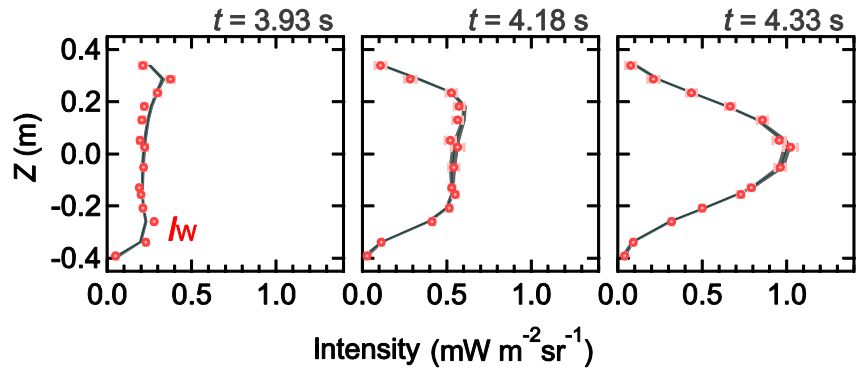

Fig.6 The LOS-integrated intensity of the tungsten ions observed at $t = $ 3.93, 4.18 and 4.33 s, with the uncertainties indicated by the error bars, which are similar to the marker size. The solid black curves show the reconstructed results of the Abel inversion.

The $I_W$ distribution has nearly symmetric profiles against $Z = 0$ m. The distribution is hollow at $t = 3.93$ s and it has a peak at $Z = 0$ m at $t = 4.33$ s. The symmetric distribution of $I_W$ suggests that this tungsten emission line comes from the inside of the LCFS, because the ergodic layer has a asymmetric structure against $Z = 0$ m while the closed magnetic flux surfaces have nearly symmetric profile (see Fig.1(a)). The temporal change of the $I_W$ distribution indicates the changes of the emission location along the minor radius $\rho$. We assume that the emissivity [mW m$^{-3}$sr$^{-1}$], which is defined as the emission power from unit volume to unit solid angle, of the tungsten emission line is a function of $\rho$, i.e., the same emissivity is assumed for the same $\rho$ positions. Under the assumption, we estimate the emissivity distribution as a function of $\rho$ by the Abel inversion of the observed LOS-integrated intensity distribution. Since the inversion is numerically unstable, we adopt a regularization method called as "Uniform penalty" [14,15]. This regularization method assumes that the emissivity distribution is a smooth function of $\rho$ and has a simple spatial structure.

The reconstructed LOS-integrated intensity distributions are shown by red solid curves in Fig.6 and estimated emissivity profiles are shown the upper panels in Fig.7. The thickness of the curves in the upper part of Fig.7 indicates the estimated uncertainty of the profile. Since the emission from the plasma center is observed with only a few LOSs around $Z = 0$ m, while that from the edge region is observed with all the LOSs, the uncertainty of the evaluated emissivity in the plasma center is larger than that in the edge region. In Fig.5 (a), the temporal evolution of the peak position of the emissivity and the emission region, which is here defined as the region with the emissivity with larger than half maximum (horizontal arrows in Fig.7), are shown.

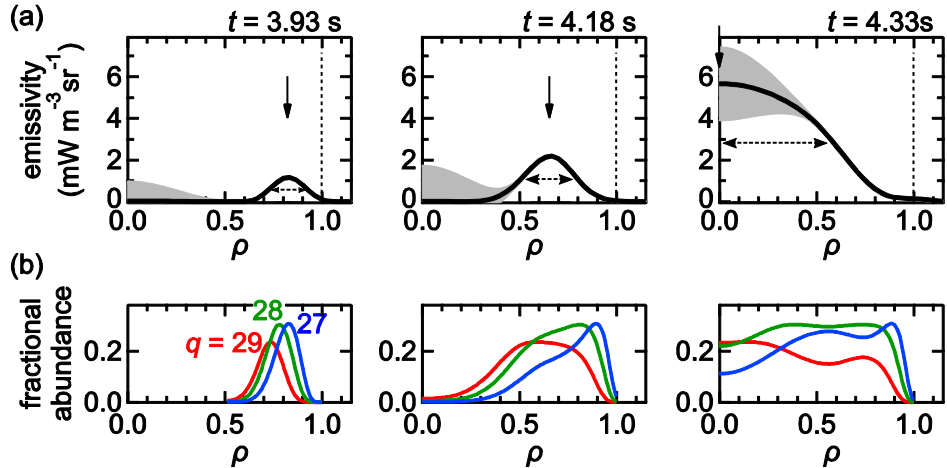

Fig.7 Radial distributions of the emissivity of the tungsten emission line estimated from the Abel inversion. The thickness of the curves shows the uncertainty of the inversion. The corresponding uncertainty of the reconstructed LOS-integrated intensity is shown by the thickness of the red curves in Fig.6. The peak positions and widths of the emissivity are indicated by the vertical and horizontal arrows, respectively.

## 4 Discussions

In Fig.5 (a), we also show the temporal evolution of $T_{Ar}^{16+}$. Although the Abel inversion analysis of $Ar^{16+}$ emission is currently unavailable, we confirmed that its intensity distribution integrated along the LOS observed by the XICS is similar to that of the tungsten line. It suggests the similar emission locations of the tungsten line and the $Ar^{16+}$ line. The difference of $T_W$ and $T_{Ar}^{16+}$ is small (less than 30 %), and both the temperatures decreases in $t > 4.0$ s. Since these ions are dominantly collided with hydrogen ions, it is suggested that the argon ions, tungsten ions, and hydrogen ions are close to in the local thermal equilibrium.

In Fig.5 (b), the temporal evolution of the position where $T_e = 0.8$ keV. The emissivity peak position is roughly coincident with the $T_e = 0.8$ keV position. The behavior of the emission profile is mainly due to the temporal change in $T_e$ profile in the plasma, i.e. in higher $T_e$ region the dominant tungsten ion charge becomes larger than the charge state of this emission line and this emission intensity there becomes smaller. Although that the total tungsten ion density distribution is unavailable, the bremsstrahlung intensity distribution suggests that it does not have large spatial gradient in the plasma.

We plot the radial distribution of the fractional abundance of $W^{27}$, $W^{28}$, and $W^{29}$ states from the radial distribution of $T_e$ and the calculated data by Sasaki et al [16] in Fig.7 by red, green, and blue curves, respectively. The emissivity distribution for $t = 3.93$ s is close to the fractional abundance profiles for $W^{27}$ and $W^{28}$ while that for $t = 4.33$s is close to the one for $W^{29}$. It suggests that this emission line is due to the tungsten ions between charge state of 27 ~ 29. The possible candidates are $W^{27+}$ [$(4d_{5/2}^5 4f_{5/2})_2 4f_{7/2}]_{13/2}$ - $[4d_{5/2}^5 4f_{5/2}^2]_{13/2}$ (the calculated wavelength is 669.3 nm [17]) and $W^{28+}$ $[4d_{5/2}^5 4f_{5/2}]_5$ - $[4d_{5/2}^5 4f_{5/2}]_6$ (the calculated wavelength is 605.57 nm [18]).

The Zeeman profile may be useful to identify its transition. When the normal Zeeman effect is assumed for the emission line, the spectral split $\delta\lambda$ is expressed as;

$$\delta\lambda = g\mu_B B \frac{\lambda_0^2}{hc} \qquad (3)$$

where $g$, $\mu_B$, $B$, c and $h$ are Landé's $g$-factor, Bohr magneton, magnetic field strength, light speed and Planck's constant, respectively. As described above, the emission of the tungsten line at $t = 4.33$ s is localized at $\rho \sim 0.0$, where the magnetic field strength there is 2.64 T. We estimated the $g$ factor from $\delta\lambda$ of the spectra observed at $t = 4.33$ s because the emission location is the center of the plasma and emission intensity is highest. The evaluated value is $0.94 \pm 0.04$. This value will be useful to make a cross-check of the future line identification by comparing with the result of perturbation theory for the Zeeman profile calculation.

## 5 Conclusion

We demonstrated a polarization-resolved and multi-LOS spectroscopy of a visible emission line of highly charged tungsten ions $\lambda_0 = 668.899$ nm in LHD plasma with 50 ms time resolution. The tungsten ion temperature was evaluated from the broadening of this emission line for the first time with the emission location which is determined from the Abel inversion. The Zeeman profile of the highly charged tungsten ions was also detected for the first time. Such a visible observation could be one candidate to measure the magnetic strength and angle as well as the ion temperature in future devices.

The local emissivity is also compared with the fractional abundance distribution of several charge state tungsten ions. From the comparison, $W^{27} \sim W^{29}$ were expected as the charge state of the emission line.

**Acknowledgement**

This work was supported by the National Institute for Fusion Science (NIFS13KLPH021 and NIFS12KOAH028).